\documentclass[preprintnumbers,amsmath,amssymb,12pt]{article}

\usepackage{graphicx}% Include figure files
\usepackage{dcolumn}% Align table columns on decimal point
\usepackage{bm}% bold math
\usepackage{graphicx}
\usepackage{amsmath,amssymb,bm}
\usepackage{graphicx,color}
\usepackage{amsmath}
\usepackage{braket}
\usepackage{amsmath,amssymb}
\usepackage{bm}
\usepackage{graphicx}
\usepackage{subfigure}
\usepackage{verbatim}
\usepackage{wrapfig}
\usepackage{ascmac}
\usepackage{makeidx}
\usepackage{bm}
\usepackage{setspace}
\usepackage{amsmath}
\usepackage{cite}
\usepackage{authblk}

\usepackage{typearea}
\typearea{12}

\begin{document}
\title{ {\Large{\textbf{Effects of entanglement on vortex dynamics in the hydrodynamic representation of quantum mechanics}}}}
{{\author[]{{Satoya Imai$^{1,2}$}}}}
\date{%
    $^1$Naturwissenschaftlich-Technische Fakult\"{a}t, Universit\"{a}t Siegen, Walter-Flex-Str. 3, D-57068 Siegen, Germany\\%
    $^2$Department of Physical Engineering, Mie University, Mie 514-8507, Japan\\[2ex]%
    \today
}

\maketitle

\noindent {\textbf{Abstract.}}
The hydrodynamic representation of quantum mechanics describes virtual flow as if a quantum system were fluid in motion.
This formulation illustrates pointlike vortices when the phase of a wavefunction becomes nonintegrable at nodal points.
We study the dynamics of such pointlike vortices in the hydrodynamic representation for a two-particle wavefunction.
In particular, we discuss how quantum entanglement influences vortex-vortex dynamics.
For this purpose, we employ the time-dependent quantum variational principle combined with the Rayleigh-Ritz method.
We analyze the vortex dynamics and establish connections with Dirac's generalized Hamiltonian formalism.

%%%%%%%%%%%%%%%%%%%%%%%%%%%%%%%%%%%%%%%%%%%%%%%%%%%%%%%%%%%%%%%%%%%%%%%%%%%%%%%%%%%%%%%%%%%%%%%%%%%%%%%%%%%%%%%%%%%%%%%%%%%%%%%%%%%%%%%%%%%%%%%%%%%%%%%%%%%%%%%%%%%%%%%%%%%%%%%%%%%%%%%%%%%%%%%%%%%%%%%%%%%%%%%%%%%%%%%%%%%%%%%%%%%

\newpage
\subsection*{1. Introduction}
There is a mathematical analogy between hydrodynamics and quantum mechanics.
The analogy virtually represents a quantum state as if it were in fluid dynamics.
The hydrodynamic representation of quantum mechanics was initially proposed by Madelung in 1926 \cite {madelung}.
Historically, these works have played a fundamental role in quantum theory and continuously fascinated the minds of physicists \cite{quantumpotential, holland, takabayasi1, takabayasi2, takabayasi3, bohm vigier, kostin1, kostin2}.

In recent years, many works have discussed these subjects from several viewpoints: quantum equilibrium hypothesis \cite{Valentini911,Valentini912}, initial-value problem \cite{2}, decomposition of the non-relativistic field velocity \cite{3}, Fisher information \cite{5,6}, quantum wave packet \cite{7,8,9}, solitary waves and nonlinear Schr\"{o}dinger equation \cite{10,11,magma}, non-Abelian fluid \cite{12,Jackiw04}, vortex dynamics in quantum trajectories \cite{18,19,20}, convective diffusion in complex hydrodynamics \cite{Tsekov12}, symmetries and conservation laws using the Noether's theorem \cite{Holland12}, quantum teleportation \cite{22}, the Klein-Gordon-Einstein equations in a weakly relativistic regime \cite{Suarez15}, the Navier-Stokes equation for viscous fluid \cite{Cordoba16}, thermo-hydrodynamic like description with Fisher information \cite{HeifetzCohen15, 21}, geometric structures of quantum theory \cite{gsqt}, propagation equation for particle probability densities \cite{propaofapr}, analytic self-similar solutions for Madelung's representation \cite {assfmr}, turbulent hydrodynamics in the Reynolds equation \cite{thre}, and Finslerian geometrization of quantum mechanics \cite{Fgqm}.

The Madelung hydrodynamic representation comes from the time-dependent Schr\"{o}dinger equation in terms of amplitude and phase of a wavefunction.
The phase factor is known to violate its integrability at nodal points, where the amplitude vanishes.
Such a nonintegrable phase can induce pointlike vortices on two-dimensional space.
Here, it is desirable to develop the hydrodynamic representation with nonintegrable phases by analogy with vortex dynamics in classical two-dimensional flow \cite{classicallamb32}.

In this paper, we study quantum vortex dynamics in the hydrodynamic representation for a two-particle wavefunction with nonintegrable phases.
In particular, our motivation is to relate vortex dynamics with entanglement, a type of non-local quantum correlation.
We purpose to explore exciting aspects of entanglement within the hydrodynamical framework.
Our idea is to introduce an ansatz for a two-particle entangled wavefunction that expresses pairs of clockwise and anticlockwise pointlike vortices.

On the other hand, the wavefunction ansatz cannot be an exact solution of the Schr\"{o}dinger equation describing such a vortex system.
To deal with this issue, we employ the time-dependent variational principle developed by Dirac \cite{Variational1, Variational2} combined with the Rayleigh-Ritz method.
This method enables us to treat pointlike vortices as collective coordinates and to obtain the equations of motion for vortex coordinates approximately.

Here, we address the non-trivial question of how entanglement influences the approximate vortex-vortex dynamics.
Our results go beyond straightforward analogies with classical hydrodynamics, at least in two respects.
First, we show that the presence of entanglement can generate highly complicated nonlinear vortex dynamics, unlike simple linear dynamics in the absence.
Second, we find that stronger entanglement can yield faster dynamics in vortex systems.
For simplicity, we will consider some restrictions on vortex coordinates.
Also, we provide a quantitative analysis of entanglement in vortex systems.
Moreover, we establish connections with Dirac's generalized Hamiltonian formalism \cite{Dirac01}.
These results may shed further light on the fundamental link between entanglement and quantum-mechanical hydrodynamics.

%%%%%%%%%%%%%%%%%%%%%%%%%%%%%%%%%%%%%%%%%%%%%%%%%%%%%%%%%%%%%%%%%%%%%%%%%%%%%%%%%%%%%%%%%%%%%%%%%%%%%%%%%%%%%%%%%%%%%%%%%%%%%%%%%%%%%%%%%%%%%%%%%%%%%%%%%%%%%%%%%%%%%%%%%%%%%%%%%%%%%%%%%%%%%%%%%%%%%%%%%%%%%%%%%%%%%%%%%%%%%%%%%%%
\subsection*{2. Hydrodynamic representation of quantum mechanics}
In this section, we describe the hydrodynamic representation of quantum mechanics.
We begin by considering the polar form of a wavefunction in position representation:
\begin{align} \label{wf}
\Psi(\boldsymbol{r}, t) = \sqrt{\rho}(\boldsymbol{r}, t) \mathrm{e}^{i S\left(\boldsymbol{r}, t\right)/\hbar}, 
\end{align}
where $\boldsymbol{r}  = (x,y,z)$.
The amplitude $ \rho (\boldsymbol{r}, t)$ is defined by
\begin{align}
\rho(\boldsymbol{r}, t)= |\Psi(\boldsymbol{r}, t)|^2,
\end{align}
and satisfies the normalization condition. 
The phase $S(\boldsymbol{r}, t)$ is defined by
\begin{align}   \label{definiyionof phase}
S(\boldsymbol{r}, t) = \frac{\hbar}{2i} \mathrm{log} \left( \frac{\Psi(\boldsymbol{r}, t)}{\Psi^*(\boldsymbol{r}, t)} \right),
\end{align}
and has the same dimension as the action.
The wavefunction (\ref{wf}) satisfies the time-dependent Schr\"{o}dinger equation in position representation
\begin{align} \label{SE}
i\hbar\frac{\partial \Psi(\boldsymbol{r}, t)}{\partial t} = -\frac{\hbar^2}{2m}\nabla^2 \Psi(\boldsymbol{r}, t) + U(\boldsymbol{r}, t)\Psi(\boldsymbol{r}, t),
\end{align}
where $\nabla = \left(\partial/\partial x, \partial/\partial y, \partial/\partial z\right)$ and $U(\boldsymbol{r}, t)$ is a time-dependent trapping potential.
Henceforth, we consider units where $\hbar= m = 1$.
Substituting the wavefunction  (\ref{wf}) into the Schr\"{o}dinger equation (\ref{SE}), we have
\begin{align} \label{ME1}
\frac{\partial \rho}{\partial t}& + \nabla\cdot(\rho{\mbox{\boldmath $u$}})  = 0, 
\\       \label{ME2} 
\frac{\partial S}{\partial t}& = - (K+Q+U),
\end{align}
where $\mbox{\boldmath $u$} \equiv \nabla S$, $K \equiv {\mbox{\boldmath $u$}^2}/{2}$, and $Q \equiv -{\nabla^2\sqrt{\rho}} /({2\sqrt{\rho})}$ are respectively called the velocity vector field, the kinetic energy, and the quantum potential \cite{quantumpotential}.
From the definition (\ref{definiyionof phase}), $\mbox{\boldmath $u$}$ is also written as
\begin{align}
\mbox{\boldmath $u$} = \nabla \left[ \frac{1}{2i} \mathrm{log} \left(\frac{\Psi}{\Psi^*} \right) \right].
\end{align}
Eq.~(\ref{ME1}) describes the conservation law for probability in quantum mechanics, whereas Eq.~(\ref{ME2}) is known as the quantum Hamilton-Jacobi equation (see \cite{holland}).
Taking the gradient of Eq.~(\ref{ME2}), we find the following nonlinear equation:
\begin{align} \label{ME3}
\frac {\partial {\mbox{\boldmath $u$}} } {\partial t} + ( {\mbox{\boldmath $u$}} \cdot \nabla) {\mbox{\boldmath $u$}} = - \nabla (Q+U),
\end{align}
where we use that $({\mbox{\boldmath $u$}}\cdot\nabla){\mbox{\boldmath $u$}} = \nabla K + (\nabla \times \mbox{\boldmath $u$} )\times{\mbox{\boldmath $u$}}$ and $\nabla \times \mbox{\boldmath $u$}=\mbox{\boldmath$0$}$.
Eqs.~(\ref{ME1}) and (\ref{ME3}), respectively, correspond to the mass conservation law and the Euler equation in classical hydrodynamics.
Both equations draw the analogy between hydrodynamics and quantum mechanics.

The hydrodynamic representation of quantum mechanics is interesting for at least four reasons.
First, it is strictly based on the position representation of the quantum state: $\Psi(\boldsymbol{r}, t) = \Braket{\boldsymbol{r}|\Psi(t)}$.
That is, even if considering the momentum representation, one cannot obtain Eqs.~(\ref{ME1}) and (\ref{ME3}).
Second, the hydrodynamic representation is mathematically equivalent to the time-dependent Schr\"{o}dinger equation.
This fact implies that there is no dissipative term violating its time-reversal invariance.
Third, the hydrodynamic representation illustrates the virtual flow describing the probability current in the position representation as if the isolated quantum system were fluid in motion.
 
Finally, the flow of the velocity vector field is irrotational, that is, the vorticity naively vanishes:
\begin{align}
{\mbox{\boldmath $\omega$}} &\equiv \nabla \times \mbox{\boldmath $u$} \nonumber \\ 
&= \nabla \times \nabla S \nonumber \\
&= \mbox{\boldmath$0$}.
\end{align}
However, this is not the case when the phase becomes nonintegrable.
Such situations arise at the position where the amplitude vanishes: $\rho = 0$.
These ideas were initially considered by Dirac in his theory of magnetic monopoles \cite{mono1,mono2}.

%%%%%%%%%%%%%%%%%%%%%%%%%%%%%%%%%%%%%%%%%%%%%%%%%%%%%%%%%%%%%%%%%%%%%%%%%%%%%%%%%%%%%%%%%%%%%%%%%%%%%%%%%%%%%%%%%%%%%%%%%%%%%%%%%%%%%%%%%%%%%%%%%%%%%%%%%%%%%%%%%%%%%%%%%%%%%%%%%%%%%%%%%%%%%%%%%%%%%%%%%%%%%%%%%%%%%%%%%%%%%%%%%%%
\subsection*{3. Wavefunction with nonintegrable phase}
In this section, we consider the hydrodynamic representation for a single-particle wavefunction with a nonintegrable phase.
It is well known that the phase becomes nonintegrable at the position where the amplitude vanishes.
In particular in two dimensions, the position is known as the nodal point, which can be regarded as a pointlike vortex.
In this paper, we attempt to provide a new theoretical formulation for the dynamics of such pointlike vortices.
However, only using the polar form of the wavefunction is insufficient to give the formulation.
To cope with this situation, we propose a wavefunction with time-dependent nodal points.
Such a wavefunction will promise to be suitable to describe the pointlike vortex moving on two-dimensional space.

Let us write a normalized wavefunction in the $xy$-plane
\begin{align} \label{WV2}
{\psi}(\boldsymbol{r}, t) = A\{x-X(t)+i\epsilon[y-Y(t)]\},
\end{align}
where $\boldsymbol{r} = (x, y)$, $\epsilon =\pm1$, and $A$ is assumed to be a normalization factor.
The amplitude of this wavefunction is given by
\begin{align} \label{amplitudevani}
|{\psi}(\boldsymbol{r}, t)|^2 = A^2 \{[x-X(t)]^2+[y-Y(t)]^2\}.
\end{align}
Then it vanishes if $\boldsymbol{r} = {\mbox{\boldmath $X$}}(t) = (X(t),Y(t))$, which is one nodal point.

Now, the concept of a pointlike vortex induced by the nonintegrable phase can be introduced into the hydrodynamic representation:
\begin{align} \label{V2}
[\nabla \times {\mbox{\boldmath $u$}}]_z = 2 \epsilon \pi \delta^2(\boldsymbol{r} - {\mbox{\boldmath $X$}}(t)).
\end{align}
Here, $\epsilon = \pm1$ denotes the sign of the pointlike vortex, where $\epsilon = -1$ $(+1)$ expresses clockwise (anticlockwise).
Eq.~(\ref{V2}) means that the pointlike vortex appears at the nodal point $\boldsymbol{r} = {\mbox{\boldmath $X$}}(t)$ in the direction of the virtual $z$-axis perpendicular to the $xy$-plane.
The pointlike vortex can move on the $xy$-plane in time but becomes at rest if ${\mbox{\boldmath $X$}}(t) = {\mbox{\boldmath $0$}}$.

%%%%%%%%%%%%%%%%%%%%%%%%%%%%%%%%%%%%%%%%%%%%%%%%%%%%%%%%%%%%%%%%%%%%%%%%%%%%%%%%%%%%%%%%%%%%%%%%%%%%%%%%%%%%%%%%%%%%%%%%%%%%%%%%%%%%%%%%%%%%%%%%%%%%%%%%%%%%%%%%%%%%%%%%%%%%%%%%%%%%%%%%%%%%%%%%%%%%%%%%%%%%%%%%%%%%%%%%%%%%%%%%%%%
\subsection*{4. Variational approach}
In this section, we discuss how to describe the vortex dynamics in the hydrodynamic representation.
First, we present an ansatz for a two-particle entangled wavefunction with nonintegrable phases.
This ansatz expresses clockwise and anticlockwise pointlike vortex pairs.
Next, we provide a powerful approximation tool for the time-dependent variational approach.
This tool enables us to obtain the equations of motion for vortex coordinates.
Finally, from more general perspectives of the variational approach, we consider other trial wavefunctions beyond the current ansatz.

\subsubsection*{4.1. Ansatz}
Let us introduce an ansatz for a two-particle entangled wavefunction
\begin{align} \label{WV22}
{\Phi}_\text{ansatz} (\boldsymbol{r}_1, \boldsymbol{r}_2, t) = N \left[\lambda\psi_1\otimes \phi_1+ (1-\lambda)\psi_2 \otimes \phi_2\right],
\end{align}
where 
\begin{align} \label{wv1}
\psi_1(\boldsymbol{r}_1, t) &= \left\{ x_1-X_1(t)+i\epsilon_1[y_1-Y_1(t)] \right\} \mathrm{e}^{-\alpha(x_1^2+y_1^2 )/2}, \\       \label{wv2}
\psi_2(\boldsymbol{r}_1, t) &= \left\{ x_1-X_1(t)+i\epsilon_2[y_1-Y_1(t)] \right\}\mathrm{e}^{-\alpha(x_1^2+y_1^2 )/2}, \\       \label{wv3}
\phi_1(\boldsymbol{r}_2, t) &= \left\{ x_2-X_2(t)+i\gamma_1[y_2-Y_2(t)] \right\}\mathrm{e}^{-\alpha(x_2^2+y_2^2 )/2}, \\      \label{wv4}
\phi_2(\boldsymbol{r}_2, t) &= \left\{ x_2-X_2(t)+i\gamma_2[y_2-Y_2(t)] \right\}\mathrm{e}^{-\alpha(x_2^2+y_2^2 )/2}.
\end{align}
The wavefunction ${\Phi}_\text{ansatz}$ defined on Hilbert space $\mathcal{H_\psi} \otimes \mathcal{H_\phi}$ is entangled if $\lambda \neq 0$.
Each of unnormalized wavefunctions, $\psi_i$, $\phi_i$, defined  on local Hilbert spaces $\mathcal{H_\psi}$, $\mathcal{H_\phi}$ is written in the similar form of the wavefunction (\ref{WV2}).
In these expressions, $N$ is the normalization factor that will be given below, and
\begin{align}
0 \leq \lambda <\frac{1}{2}
\end{align}
is the entanglement parameter, which we will later explain why $\lambda$ does not become $1/2$.
In addition, $\epsilon_i, \, \gamma_i =\pm 1$ denote the signs of the vortices.
To distinguish two wavefunctions, $\psi_1$, $\psi_2$ (or $\phi_1$, $\phi_2$), we impose
\begin{align}  \label{conditionepgam}
\epsilon_1\epsilon_2 = \gamma_1\gamma_2 = -1.
\end{align}
This condition implies that the rotational directions between two vortices are opposite.
Therefore, $\psi_1$, $\psi_2$ (or $\phi_1$, $\phi_2$) are mutually identified by the different vortex signs.
Moreover, to deal with the regularization of divergent integrals, we introduce a positive real constant factor of $\alpha$ in the Gaussian.
This procedure makes the wavefunction normalizable.
The key idea in this paper is the ansatz (\ref{WV22}) that is the two-particle entangled wavefunction with nonintegrable phases describing clockwise and anticlockwise pointlike vortex pairs.
Intuitively, this entanglement in vortex systems may look like two-particle entanglement in spin $1/2$ systems, in the sense that clockwise and anticlockwise pointlike vortices can be analogous with quantum spin up and down.

Using the normalization condition
\begin{align}
\int_{-\infty}^{\infty} \limits \int_{-\infty}^{\infty} \limits d^2{\boldsymbol{r}}_1\, d^2{\boldsymbol{r}}_2 \; |{\Phi}_\text{ansatz} (\boldsymbol{r}_1, \boldsymbol{r}_2, t) |^2 =1,
\end{align}
we obtain
\begin{equation} \label{normali}
N = \frac{\alpha/\pi} {\sqrt {\Lambda \left(\frac{1}{\alpha} + X_1^2 + Y_1^2 \right) \left(\frac{1}{\alpha} + X_2^2 + Y_2^2 \right)
        +\Upsilon \left[ \left(X_1^2 - Y_1^2 \right) \left(X_2^2 - Y_2^2 \right) + \mu X_1Y_1X_2Y_2 \right] } },
\end{equation}
where
\begin{align}
\Lambda &= \lambda^2 + (1-\lambda)^2,
\\
\Upsilon &= 2\lambda(1-\lambda),
\\
\mu &= -(\epsilon_1 - \epsilon_2)(\gamma_1-\gamma_2)=\pm 4.
\end{align}
The normalization factor $N$ is not a constant because each of the vortex variables, $X_i$, $Y_i$, depends on time.

Note that there is no unique ansatz for a wavefunction describing a vortex system.
As long as the amplitude vanishes at nodal points, the ansatz would be proposed.
Also, some restrictions on three-dimensional space would be considered.
On the other hand, we desire to discuss new findings without handling such complicated situations.
To this end, we require a simple ansatz that can be used for analytical calculations.

The wavefunction ansatz (\ref{WV22}) is a type of non-Gaussian entangled states in continuous-variable systems.
Non-Gaussian entangled states have recently been required for many quantum information processing tasks, e.g., universal quantum computation \cite{quantumcomuter1999}, teleportation \cite{telepo2000}, and violations of Bell inequalities \cite{Bellinequality11, Bellinequality22}.
Such non-Gaussian states can be expected to possess new quantum-mechanical powers beyond Gaussian states.
In this sense, non-Gaussianity can be considered as a physical resource \cite{QRT11, QRT22}.
In particular, the ansatz (\ref{WV22}) may seem to have a form similar to a special case of the well-known Laguerre-Gaussian wavefunctions that have been experimentally realized and demonstrates EPR steering \cite{EPRteering2014}.
In addition, in subsection 5.3, we will show that the von Neumann entanglement entropy of the ansatz does not vanish.
This signature of entanglement can imply violations of inequalities from the assumptions of locality and realism.
On the other hand, it would be challenging to develop our framework from these perspectives of non-Gaussianity and non-locality.
This is because the Madelung hydrodynamic representation depends only on the position representation and describes the virtual flow as an ideal physics.

%%%%%%%%%%%%%%%%%%%%%%%%%%%%%%%%%%%%%%%%%%%%%%%%%%%%%%%%%%%%%%%%%%%%%%%%%%%%%%%%%%%%%%%%%%%%%%%%%%%%%%%%%%%%%%
\subsubsection*{4.2. Time-dependent variational approach}
Here, we describe a theoretical tool for developing vortex dynamics based on the ansatz.
We stress that the ansatz cannot be an exact solution of the two-body time-dependent Schr\"{o}dinger equation.
Nevertheless, we hope that the ansatz is a clear approximate solution that adequately represents the quantum model of the vortex system.
To address this situation, we employ the time-dependent variational principle for quantum mechanics developed by Dirac \cite{Variational1} (see \cite{Variational2}).
This method enables us to propose the ansatz as a trial wavefunction and to obtain an approximate solution.

Let us begin by writing the action
\begin{align} \label{Action}
I &= \ \int_{t_i}^{t_f} dt \; L,
\end{align}
where $t_f$, $t_i$ are respectively the final and initial times.
In the present case, the Lagrangian is given by
\begin{align}       \label{Lagrangian2}
L&=\int_{-\infty}^{\infty} \limits \int_{-\infty}^{\infty} \limits d^2{\boldsymbol{r}}_1\, d^2{\boldsymbol{r}}_2 \; \Psi ^*(\boldsymbol{r}_1, \boldsymbol{r}_2 ,t)\left(i\frac{\partial}{\partial t}-H\right)\Psi(\boldsymbol{r}_1, \boldsymbol{r}_2 ,t),
\end{align}
where $H= - {\nabla_1}^2/2 - {\nabla_2}^2/2 + U(\boldsymbol{r}_1, \boldsymbol{r}_2 ,t)$, $U(\boldsymbol{r}_1, \boldsymbol{r}_2 ,t)$ is a time-dependent trapping potential, and $\nabla_i = \left(\partial/\partial x_i, \partial/\partial y_i\right)$, $i=1,2$.
It is well known that the time-dependent variational principle for the Schr\"{o}dinger equation automatically satisfies the normalization constraint and the time boundary conditions.
This is because we here do not consider the conditions.

Let us employ the variational principle combined with the Rayleigh-Ritz method.
We adopt the wavefunction (\ref{WV22}) with the normalization factor (\ref{normali}) as a trial wavefunction.
Substituting it into the Lagrangian, and performing the integral with respect to ${\boldsymbol{r}}_1$, ${\boldsymbol{r}}_2$ in position space, we obtain
\begin{align} \label{L2}
L = &\frac{ E \left(\frac {1} {\alpha} + X_2^2 + Y_2^2\right) ({\dot{X}}_1Y_1 - X_1{\dot{Y}}_1)
                             + \Gamma \left(\frac {1} {\alpha} + X_1^2 + Y_1^2\right) ({\dot{X}}_2Y_2 - X_2{\dot{Y}}_2) }
                           {\Lambda \left(\frac{1}{\alpha} + X_1^2 + Y_1^2 \right) \left(\frac{1}{\alpha} + X_2^2 + Y_2^2 \right)
                                          + \Upsilon \left[(X_1^2 - Y_1^2)(X_2^2 - Y_2^2) + \mu X_1Y_1X_2Y_2 \right]} \nonumber\\
                     &-\frac{\Lambda}{2} \frac{\frac{2}{\alpha} + X_1^2 + Y_1^2 + X_2^2 + Y_2^2 }
                           {\Lambda \left(\frac{1}{\alpha} + X_1^2 + Y_1^2 \right) \left(\frac{1}{\alpha} + X_2^2 + Y_2^2 \right)
                                          + \Upsilon \left[(X_1^2 - Y_1^2)(X_2^2 - Y_2^2) + \mu X_1Y_1X_2Y_2 \right]} \nonumber\\
                     &-\bar{U}(X_1,Y_1,X_2,Y_2,t),
\end{align}
where an irrelevant additive constant has not been included, and the over-dot denotes the time derivative.
In this expression, we define
\begin{align}
\bar{U}(X_1,Y_1,X_2,Y_2,t) &= \int_{-\infty}^{\infty} \limits \int_{-\infty}^{\infty} \limits d^2{\boldsymbol{r}}_1\, d^2{\boldsymbol{r}}_2 \;  {\Phi}_\text{ansatz}^* (\boldsymbol{r}_1, \boldsymbol{r}_2, t) U(\boldsymbol{r}_1, \boldsymbol{r}_2 ,t) {\Phi}_\text{ansatz} (\boldsymbol{r}_1, \boldsymbol{r}_2, t),
\\
E &=\lambda^2\epsilon_1\,+\, (1-\lambda)^2\epsilon_2,
\\
\Gamma &=\lambda^2\gamma_1 \,+\, (1-\lambda)^2\gamma_2.
\end{align}

This time-dependent variational approach derives the reduced Lagrangian for the vortex variables, $L(X_1, Y_1, X_2,$ $Y_2, \dot{X}_1, \dot{Y}_1, \dot{X}_2, \dot{Y}_2)$, from the Lagrangian (\ref{Lagrangian2}), as an effective description.
The advantage of this approach is that by performing the integral in position space, we can focus only on the degrees of freedom related to the vortex system and eliminate the other degrees of freedom.
To take this advantage, we can treat pointlike vortices as collective coordinates.
Then we can turn the vortex coordinates into the generalized coordinates in the phase space.

It is convenient to assume that the vortex dynamics can be localized to a time-dependent trapping potential.
To do so, we assume that $\bar{U}$ is approximated to a constant and therefore can be ignored.
In addition, we require that the value of $\alpha$ be large so that the Gaussian wave packet does not spread.

%%%%%%%%%%%%%%%%%%%%%%%%%%%%%%%%%%%%%%%%%%%%%%%%%%%%%%%%%%%%%%%%%%%%%%%%%%%%%%%%%%%%%%%%%%%%%%%%%%%%%%%%%%%%%%
\subsubsection*{4.3. Other trial wavefunctions}
Here, apart from two-particle systems, we discuss the time-dependent variational approach based on other trial wavefunctions.
Although this subsection is not the main part of this paper, we hope that such a discussion will be helpful for further theoretical studies using variational methods.

%%%%%%%%%%%%%%%%%%%%%%%%%%%%%%%%%%%%%%%%%%%
First, let us write the following ansatz for a single-particle wavefunction with nonintegrable phases in the $xy$-plane:
\begin{align} \label{wvnnns}
\phi_n(\boldsymbol{r}, t) = N\{x-X_1(t)+i\epsilon_1 [y-Y_1(t)]\}^{k_1} \times \cdots \times \{x-X_n(t)+i\epsilon_n [y-Y_n(t)]\}^{k_n} \mathrm{e}^{-\alpha(x^2+y^2 )/2},
\end{align}
where $\boldsymbol{r}=(x,y)$, $\epsilon_i =\pm 1$ denote the signs of the vortices, $\alpha$ is a positive real constant, and $N$ is a normalization factor.
The amplitude of this wavefunction vanishes at $\boldsymbol{r} = {\mbox{\boldmath $X$}}_i(t) = (X_i(t),Y_i(t))$ for $i=1,\ldots, n$.
Then this single-particle wavefunction has the $n$ nodal points, which can induce the $n$ pointlike vortices:
\begin{align}
[\nabla \times {\mbox{\boldmath $u$}}]_z = 2\pi k_1 \epsilon_1\delta^2(\boldsymbol{r} - {\mbox{\boldmath $X$}}_1(t)) + \cdots + 2\pi k_n \epsilon_n \delta^2(\boldsymbol{r}- {\mbox{\boldmath $X$}}_n(t)).
\end{align}
Here, the natural numbers $k_i$ express the strength of the pointlike vortices.
The constants $k_i \epsilon_i$ are referred to as the vortex charges.

In principle, the time-dependent variational approach allows us to derive the reduced Lagrangian with these vortex variables from the substitution of the wavefunction ansatz into the Lagrangian given by
\begin{align} 
L=\int_{-\infty}^{\infty} d^2{\boldsymbol{r}} \; \phi^*_n(\boldsymbol{r}, t)\left(i\frac{\partial}{\partial t}-H\right)\phi_n(\boldsymbol{r}, t),
\end{align} 
where $H= -{{\nabla}^2}/{2} + U(\boldsymbol{r} ,t)$.
On the other hand, to calculate the integral in two-dimensional space and to obtain straightforward analytical expressions, we would require fewer parameters.
For instance, let us consider that $k_1=k$, $k_2=1$, $k_{i\neq1,2}=0$, ${\mbox{\boldmath $X$}}_1(t)={\mbox{\boldmath $0$}}$, and ${\mbox{\boldmath $X$}}_2(t) = {\mbox{\boldmath $X$}}(t)$.
In this setting, the single-particle wavefunction ansatz has the two nodal points, where one is at rest, and the other can move in time.
Since the normalization factor $N$ is now calculated by
\begin{align} 
N = \sqrt{ \frac {\alpha^{k+2}/(\pi k!)} {k+1+\alpha \left(X^2+Y^2 \right)}},
\end{align}
we have the reduced Lagrangian
\begin{align} \label{L1}
L = \frac{\epsilon_2\alpha(\dot{X}Y-X\dot{Y})} {k+1+\alpha(X^2+Y^2 )} -\frac{c\alpha} {k+1+\alpha(X^2+Y^2 )},
\end{align}
where an irrelevant additive constant has not been included, $c = \epsilon_1\epsilon_2k- ({k+1})/{2}$ is a constant, and we assume that the time-dependent trapping potential is approximated to a constant. 

%%%%%%%%%%%%%%%%%%%%%%%%%%%%%%%%%%%%%%%%%%%
Next, let us extend two-particle systems with pointlike vortices to three-particle systems.
In particular, we are interested in a genuine three-particle entangled wavefunction, which is neither fully separable nor biseparable (see \cite{otfried}).
As examples of such states, we can introduce the GHZ-like state and W-like state
\begin{align}
\psi_{\textrm{GHZ}} &= N \left\{ \lambda_1 \psi_+^{\otimes 3} + \lambda_2 \psi_-^{\otimes 3} \right\}, \\
\psi_{\textrm{W}} &= N \left\{\lambda_1 \psi_+ \otimes \psi_-^{\otimes 2} + \lambda_2 \psi_- \otimes \psi_+ \otimes \psi_- + \lambda_3 \psi_-^{\otimes 2} \otimes \psi_+ \right\},
\end{align}
where $\lambda_i$ are positive entanglement parameters satisfying $\sum_i \lambda_i = 1$, $N$ is a normalization factor, and unnormalized wavefunctions, $\psi_+$ and $\psi_-$, are given by
\begin{align} \label{wvll1}
\psi_+(\boldsymbol{r}_i, t) &= \left\{ x_i-X_i(t)+i[y_i-Y_i(t)] \right\}\mathrm{e}^{-\alpha(x_i^2+y_i^2 )/2}, \\
\psi_-(\boldsymbol{r}_i, t) &= \left\{ x_i-X_i(t)-i[y_i-Y_i(t)] \right\}\mathrm{e}^{-\alpha(x_i^2+y_i^2 )/2}.
\end{align}
By using the variational approach with these trial wavefunctions, one can study the dynamics of three pointlike vortices.
It would be interesting to analyze the physical differences in three-particle entanglement effects on the two types of vortex dynamics and to compare them with the two-particle case.
Furthermore, one can generalize to genuine multi-particle entangled states in vortex systems.

%%%%%%%%%%%%%%%%%%%%%%%%%%%%%%%%%%%%%%%%%%%
Finally, let us give a general explanation of the time-dependent quamtum variational approach.
The variational method is an approximate way to solve the time-dependent Schr\"{o}dinger equation.
Rather than considering the exact analytical solution, one offers a so-called trial wavefunction, which would be the most appropriate solution for the variations.
Such a trial wavefunction essentially represents a physical model and contains several parameters.
The task is to compute the best approximation of the analytical solution for these parameters.
The main benefit of this variational approach is that it can be applied to any form of Hamiltonian.
For this purpose, it is necessary to choose the trial wavefunction that has a computationally convenient and physically acceptable form.

A typical example is the Gaussian trial wavefunction
\begin{align} \label{wavefunctionharmonicgaussian}
\Psi_G \, (x, t) = N \, \mathrm{exp} \left\{ {-\frac{\left[x-X(t) \right]^2 } {2\left[\Sigma^{2}(t) +iG(t) \right]} } +iP(t)\left[x-X(t) \right] \right\},
\end{align}
where $N$ is a normalization factor, and $X(t)$, $P(t)$, $\Sigma(t)$, and $G(t)$ are the time-dependent coordinate variables in the one-dimensional position space.
This function is called the Gaussian ansatz for the wavefunction.
The first and second terms in the exponent respectively describe localization and propagation of the plane wave.
Since this trial wavefunction has the quadratic form, one can straightforwardly perform the analytical calculation.
We remark that the Gaussian ansatz for the density matrix is known to be used for the variational principle for the Liouville-von Neumann equation \cite{eboli}.

Let us consider the case of the harmonic oscillator.
Then the Gaussian ansatz becomes the exact solution of the time-dependent Schr\"{o}dinger equation.
Here, solving the Schr\"{o}dinger equation means knowing how the time-dependent coordinate variables evolve in time.
One way to know that is to directly substitute the Gaussian ansatz into the time-dependent Schr\"{o}dinger equation.
This way leads to the equations for the coordinate variables.
Another way is to apply the Gaussian ansatz as a trial function to the time-dependent variational approach.
This way derives the reduced Lagrangian with coordinate variables from the Lagrangian given by
\begin{align}
L = \int_{-\infty}^{\infty} dx \; \Psi^*_G \, (x,t)\left(i\frac{\partial}{\partial t} +\frac{1}{2}\frac{\partial^2}{\partial x^2} - \frac{1}{2}x^2 \right)\Psi_G\,(x, t).
\end{align}
Then, from the Euler-Lagrange equations of the reduced Lagrangian, one can know the time evolution of coordinate variables.
Either way, the same solution can be obtained.

However, if a given Hamiltonian contains an anharmonic term, for example, $x^4$, the situation changes.
This is because there is no exact solution to the Schr\"{o}dinger equation.
In this case, the first way cannot be used, but the second way (the variational approach) can be used; by applying the Gaussian ansatz to the time-dependent variational approach, one can likewise have the reduced Lagrangian and find its approximate solution from the Euler-Lagrange equations.

%%%%%%%%%%%%%%%%%%%%%%%%%%%%%%%%%%%%%%%%%%%%%%%%%%%%%%%%%%%%%%%%%%%%%%%%%%%%%%%%%%%%%%%%%%%%%%%%%%%%%%%%%%%%%%%%%%%
\subsection*{5. Entanglement and vortex dynamics}
In this section, we study how entanglement influences the approximate vortex-vortex dynamics.
We focus on the terms $\Lambda$, $E$, $\Gamma$, and $\Upsilon$ that can be characterized by the entanglement parameter $\lambda$.
Our findings highlight aspects of entanglement effects.

\subsubsection*{5.1. $\lambda$ is equal to $1/2$}
Let us see a relation between the entanglement parameter and the kinetic term in the reduced Lagrangian (\ref{L2}).
If $\lambda = 1/2$ [i.e., $E, \,\Gamma = 0$], then the kinetic terms naively vanish.
Now, a question arises: what this vanishing means physically?
Making a discussion about this particular quantum case is essential for at least two implications.

The first point is from the time-dependent variational principle for the Schr\"{o}dinger equation.
We begin by writing the complex conjugate of the wavefunction (\ref{WV22}):
\begin{align}
    {\Phi}^*_\text{ansatz} = N \left[\lambda\psi_1^* \otimes \phi_1^* + (1-\lambda)\psi_2^* \otimes \phi_2^*\right].
\end{align}
Notice that the condition (\ref{conditionepgam}), which is $\epsilon_1\epsilon_2 = \gamma_1\gamma_2 = -1$, leads to the fact that $\psi_1=\psi_2^*$ and $\phi_1= \phi_2^*$.
The complex conjugate is thus rewritten as
\begin{align}
{\Phi}^*_\text{ansatz} = N \left[\lambda\psi_2 \otimes \phi_2 + (1-\lambda)\psi_1 \otimes \phi_1 \right].
\end{align}
If $\lambda=1/2$, the complex conjugate is equal to the wavefunction (\ref{WV22}):
\begin{align} \label{realvauewafn}
    {\Phi}^*_\text{ansatz} = {\Phi}_\text{ansatz}.
\end{align}
In the scheme of the time-dependent variational principle for the Schr\"{o}dinger equation, a wavefunction and its complex conjugate must be independent of each other.
In order to distinguish both functions, the wavefunction must be a complex value.
If the wavefunction is a real value, the time-dependent variational principle does not work correctly.
Thus Eq.~(\ref{realvauewafn}) tells us that the ansatz (\ref{WV22}) will be inappropriate for a trial wavefunction if $\lambda = 1/2$.
This is why we avoided $\lambda = 1/2$ in the previous section.

The second point is from the quantification of pure two-particle entanglement.
As a measure of entanglement, consider the von Neumann entropy of partial systems.
We will later show that the von Neumann entanglement entropy of the ansatz has the largest value at $\lambda = 1/2$.
Hence, we can conclude that the time-dependent variational principle for the Schr\"{o}dinger equation is not available when quantum states in vortex systems are mostly entangled.

%%%%%%%%%%%%%%%%%%%%%%%%%%%%%%%%%%%%%%%%%%%%%%%%%%%%%%%%%%%%%%%%%%%%%%%%%%%%%%%%%%%%%%%%%%%%%%%%%%%%%%%%%%%%%%%%%%%
\subsubsection*{5.2. Nonlinear effects}
Let us discuss the effects of entanglement on the vortex dynamics.
In the first place, the terminology of vortex dynamics here means the equations of motion for the vortex coordinates $X_1$, $Y_1$, $X_2$, and $Y_2$.
The equations of motion are obtained by using the Euler-Lagrange equations calculated from the reduced Lagrangian.
Indeed, if the wavefunction ansatz (\ref{WV22}) is entangled [i.e., $\lambda \neq 0$], we can find that the Euler-Lagrange equations become highly complicated nonlinear.
On the other hand, if the ansatz is not entangled [i.e., $\lambda=0$], the Euler-Lagrange equations turn out to be simply linear, especially to behave like a harmonic oscillator.
Hence, we can naturally indicate that entanglement can generate nonlinear effects on vortex dynamics.

Another way to understand such nonlinear effects is to use the Hamiltonian formalism.
With the help of the Legendre transformation, the reduced Hamiltonian for the vortex coordinates is written as
\begin{align}  \label{hamilchap8}
H &= \dot{X}_1{p_{X_1}} + \dot{Y}_1{p_{Y_1}} + {\dot{X}}_2 {p_{X_2}} + {\dot{Y}}_2 {p_{Y_2}} - L 
\nonumber\\
&=  \frac{\Lambda}{2} \frac{\frac{2}{\alpha} + X_1^2 + Y_1^2 + X_2^2 + Y_2^2 }{ {\Lambda \left(\frac{1}{\alpha} + X_1^2 + Y_1^2 \right) \left(\frac{1}{\alpha} + X_2^2 + Y_2^2 \right)
        +\Upsilon \left[ \left(X_1^2 - Y_1^2 \right) \left(X_2^2 - Y_2^2 \right) + \mu X_1Y_1X_2Y_2 \right] } },
\end{align}
where ${p_{X_1}}$, ${p_{Y_1}}$, ${p_{X_2}}$, and ${p_{Y_2}}$ are the canonical momenta.
Setting the angular momenta as $s_1 = X_1 {p_{Y_1}} -Y_1 {p_{X_1}}$ and $s_2 = X_2 {p_{Y_2}} -Y_2 {p_{X_2}}$, we have
\begin{align} \label{NH}
H = \frac{\alpha \Lambda}{2} \left(E s_1 + \Gamma s_2 -2g s_1 s_2 \right),
\end{align}
where $g$ is a position-dependent coupling coefficient defined by 
\begin{align}
&g(X_1,Y_1,X_2,Y_2,\lambda)
\nonumber\\
&= - \frac{{\Lambda \left(\frac{1}{\alpha} + X_1^2 + Y_1^2 \right) \left(\frac{1}{\alpha} + X_2^2 + Y_2^2 \right)
        +\Upsilon \left[ \left(X_1^2 - Y_1^2 \right) \left(X_2^2 - Y_2^2 \right) + \mu X_1Y_1X_2Y_2 \right] } }{E\Gamma  \left(X_1^2 + Y_1^2 \right) \left(X_2^2 + Y_2^2 \right)}.
\end{align}
The angular momenta $s_1$ and $s_2$ are the canonical variables.
They can be respectively conjugate to the phases $\theta_1 = {\mathrm {tan}}^{-1} (Y_1/X_1)$ and $\theta_2 = {\mathrm {tan}}^{-1} (Y_2/X_2)$.
It follows that the transformation to the angular momenta is the canonical transformation, not a simple change of variables.
More specifically, there is a function $f(s_1,\theta_1 s_2,\theta_2,\lambda)$ that can be equal to $g(X_1,Y_1,X_2,Y_2,\lambda)$, although it cannot be explicitly represented.

The reduced Hamiltonian (\ref{NH}) is in analogy with the Hamiltonian in quantum spin models.
In particular, when $g<0$ (or $g>0$), i.e., $\epsilon_1 = + \gamma_1$ (or $\epsilon_1 = - \gamma_1 $), the quantum vortex system looks like to be characterized as the Ferro-coupling (Antiferro-coupling).
However, in the absence of entanglement, the coupling term, $-2gs_1 s_2$, is ignored because it becomes independent of the vortex coordinates.
There, the angular momenta become invariant in time.
This result is consistent with the harmonically oscillating vortex dynamics that we previously described.
Hence, we can suggest that the presence of the entanglement parameter strongly complicates the vortex dynamics, resulting in the nonlinear effects.

Moreover, let us discuss the vortex dynamics in terms of the difference in the vortex signs.
As clearly as possible, we perform the transformation from $(X_2,Y_2 )$ to $(-X_1,-Y_1 )$.
The reduced Lagrangian (\ref{L2}) is thus written as 
\begin{align} \label{NL}
L = \frac{\left(\frac {1} {\alpha} + X^2 + Y^2\right) 
                            \left[ (E+\Gamma)({\dot{X}}Y - X{\dot{Y}})-\Lambda \right] }
                           {\Lambda \left(\frac{1}{\alpha} + X^2 + Y^2 \right)^2 
                           + \Upsilon \left[ \left(X^2 - Y^2\right)^2 + \mu X^2Y^2 \right]},
\end{align}
where we take out the notation of subscript of $(X_1,Y_1 )$.
Now we focus on a relation between the terms $E$ and $\Gamma$, where $E =\lambda^2\epsilon_1\,+\, (1-\lambda)^2\epsilon_2$ and $\Gamma =\lambda^2\gamma_1 \,+\, (1-\lambda)^2\gamma_2$.
When $E = -\Gamma$ [i.e., $\epsilon_1 = -\gamma_1$], the kinetic term naively vanishes.
This vanishing means that the vortex system becomes static.
Note here that the wavefunction ansatz (\ref{WV22}) remains complex unless $\lambda \to 1/2$, unlike Eq.~(\ref{realvauewafn}).
On the other hand, when $E = +\Gamma$ [i.e., $\epsilon_1 = +\gamma_1$], the vortex system is no longer static, and the Euler-Lagrange equations can be nonlinear.
That is, whether the vortex dynamics behaves the nonlinear motion or not can depend on the signs of the vortices.
This result also comes from the sort of entanglement effects.
We can conclude that the vortex dynamics can be involved with not only the entanglement parameter but also the vortex signs.

%%%%%%%%%%%%%%%%%%%%%%%%%%%%%%%%%%%%%%%%%%%%%%%%%%%%%%%%%%%%%%%%%%%%%%%%%%%%%%%%%%%%%%%%%%%%%%%%%%%%%%%%%%%%%%%%%%%
\subsubsection*{5.3. Quantitative analysis of entanglement in vortex systems}
Let us study entanglement of the wavefunction ansatz (\ref{WV22}) in vortex systems from quantitative viewpoints.
For this purpose, we quantify how much the ansatz is entangled by using the von Neumann entropy: for any quantum state $\rho$,
\begin{align}
         S(\rho) =  -\mathrm{tr} \, (\rho \log{\rho}).
\end{align}
The von Neumann entropy formally corresponds to the Shannon entropy in classical information theory, which has a clear operational interpretation.
In general, for a pure two-particle state $\Ket{\psi}_{AB}$, an entanglement measure $E(\psi_{AB})$ is given by the von Neumann entropy of the reduced density matrices \cite{Braunstein, Adesso}:
\begin{align}  \label{secondequal}
    E(\psi_{AB}) = S(\rho_A) = S(\rho_B),
\end{align}
where $\rho_A = \mathrm{tr}_B \, (\Ket{\psi}\Bra{\psi}_{AB})$ and $\rho_B = \mathrm{tr}_A \, (\Ket{\psi}\Bra{\psi}_{AB})$ are respectively the reduced density matrices on the subsystems.
The von Neumann entanglement entropy naturally satisfies the following characteristic properties: (i) vanishing on non-entangled states (ii) invariance under local unitary transformations (iii) additivity (iv) monotonicity under LOCC.
Note that the second equality on the formula (\ref{secondequal}) comes from the triangle inequality showed by Araki and Lieb.

We adapt the von Neumann entanglement entropy as the entanglement measure of the wavefunction ansatz in vortex systems.
Let us begin by defining the density matrix of the ansatz
\begin{align}
\rho_\text{ansatz} = \ket{{\Phi}_\text{ansatz}} \bra{{\Phi}_\text{ansatz}},
\end{align}
where $\ket{{\Phi}_\text{ansatz}} = N \left[\lambda \Ket{\psi_1} \otimes \Ket{\phi_1} + (1-\lambda) \Ket{\psi_2} \otimes \Ket{\phi_2}\right]$ becomes the wavefunction (\ref{WV22}) when the position representation is considered.
Then we have the reduced density matrix
\begin{align} \label{rppp1}
\rho_\psi &= \mathrm{tr}_{\phi}\, (\rho_\text{ansatz})
\nonumber\\
&= N^2\left[a^{\prime} \lambda^2 \Ket{\psi_1} \Bra{\psi_1} + a^{\prime} (1-\lambda)^2 \Ket{\psi_2} \Bra{\psi_2} +  \lambda(1-\lambda) \left( b^{\prime} \Ket{\psi_1} \Bra{\psi_2} + (b^{\prime})^\ast \Ket{\psi_2} \Bra{\psi_1} \right) \right],
\end{align}
where $a^{\prime}=a_1^{\prime}=a_2^{\prime}$ and 
\begin{align}
a_1^{\prime}   &= \Braket{\phi_1|\phi_1} = \frac{\pi}{\alpha} \left(\frac{1}{\alpha} + X_2^2 + Y_2^2\right),
\\
a_2^{\prime}   &= \Braket{\phi_2|\phi_2} = \frac{\pi}{\alpha} \left(\frac{1}{\alpha} + X_2^2 + Y_2^2\right),
\\
b^{\prime}     &= \Braket{\phi_2|\phi_1} = \frac{\pi}{\alpha} \left[X_2^2 - Y_2^2 +i(\gamma_1-\gamma_2) X_2Y_2\right],
\\
(b^{\prime})^\ast     &= \Braket{\phi_1|\phi_2} = \frac{\pi}{\alpha} \left[X_2^2 - Y_2^2 -i(\gamma_1-\gamma_2) X_2Y_2\right].
\end{align}
Here we introduce $\ket{e_1}$ and $\ket{e_2}$ as follows:
\begin{align}
    \ket{e_1} &= {\frac{1}{\sqrt{a}}} \Ket{\psi_1},\\
    \ket{e_2} &= {\frac{\sqrt{a}}{\sqrt{c}}} \Ket{\psi_2} - \frac{b^\ast}{\sqrt{ac}} \Ket{\psi_1},
\end{align}
where $a=a_1=a_2$ and 
\begin{align}
a_1   &= \Braket{\psi_1|\psi_1} = \frac{\pi}{\alpha} \left(\frac{1}{\alpha} + X_1^2 + Y_1^2\right),
\\
a_2   &= \Braket{\psi_2|\psi_2} = \frac{\pi}{\alpha} \left(\frac{1}{\alpha} + X_1^2 + Y_1^2\right),
\\
b     &= \Braket{\psi_2|\psi_1} = \frac{\pi}{\alpha} \left[X_1^2 - Y_1^2 +i(\epsilon_1-\epsilon_2) X_1Y_1\right],
\\
b^\ast     &= \Braket{\psi_1|\psi_2} = \frac{\pi}{\alpha} \left[X_1^2 - Y_1^2 -i(\epsilon_1-\epsilon_2) X_1Y_1\right].
\end{align}
Notice that $\Braket{e_1|e_1} = \Braket{e_2|e_2}=1$ and $\Braket{e_1|e_2}=\Braket{e_2|e_1}=0$.
The reduced density matrix is thus rewritten as
\begin{align} 
\rho_\psi &=N^2\Biggl\{\left[a a^{\prime} \lambda^2 + \frac{a^{\prime} |b|^2}{a}(1-\lambda)^2+
(bb^{\prime}+b^\ast (b^{\prime})^\ast)\lambda(1-\lambda)\right] \Ket{e_1} \Bra{e_1} + \frac{a^{\prime} c}{a}(1-\lambda)^2 \Ket{e_2} \Bra{e_2}
\nonumber\\
&+\sqrt{c}\left[\frac{a^{\prime} b^\ast}{a} (1-\lambda)^2 + b^{\prime}\lambda(1-\lambda)\right] \Ket{e_1} \Bra{e_2}
+ \sqrt{c}\left[\frac{a^{\prime} b}{a} (1-\lambda)^2 + (b^{\prime})^\ast \lambda(1-\lambda) \right] \Ket{e_2} \Bra{e_1}\Biggr\},
\end{align}
where $c=a^2-|b|^2$.
Using the inverse unitary diagonalization, we have the von Neumann entanglement entropy
\begin{align} \label{En}
E({\Phi}_\text{ansatz}) &= S(\rho_\psi)
\nonumber\\
&=  -\sum_{p_i=p_{\pm}}{p_i\log p_i},
\end{align}
where eigenvalues $p_{\pm}$ are given by
\begin{align}
p_{\pm} = \frac{1\pm\sqrt{1-4N^2\lambda^2(1-\lambda)^2cc^{\prime}}}{2},
\end{align}
and $c^{\prime}=(a^{\prime})^2-|b^{\prime}|^2$.
Calculating the $\lambda$-derivation of the entanglement entropy
\begin{align} \label{EnD}
\frac{\partial E({\Phi}_\text{ansatz})} {\partial \lambda} =0,
\end{align}
we immediately obtain the condition $\lambda = 1/2$.
Therefore, we find that the entanglement entropy has the largest value at the mostly entangled state.
As mentioned in subsection 5.1, the ansatz (\ref{WV22}) is inappropriate for a trial wavefunction in this case.
Evidently, $S(\rho_\psi) = S(\rho_\phi)$ holds because $p_{\pm}$ are symmetric under exchange of $(X_1,Y_1)$ and $(X_2,Y_2)$.
Note that one can address the analysis of the entanglement measure beyond this result by approaching the issue that $E({\Phi}_\text{ansatz})$ cannot increase under LOCC.

Furthermore, we develop a quantitative analysis of entanglement.
Let us define a continuous-variable version of concurrence by analogy with discrete-variable cases \cite{Wootters, arXivBP}: for the pure two-particle ansatz $\ket{{\Phi}_\text{ansatz}}$, 
\begin{align}
    C({\Phi}_\text{ansatz}) &=\sqrt{2[1-\mathrm{tr} \, (\rho_{\psi}^2)]} = \sqrt{2[1-\mathrm{tr} \, (\rho_{\phi}^2)]}.
\end{align}
Note that the concurrence is a dimension-dependent factor, and in continuous-variable systems, it can exceed $1$.
Now, we can expect that $C({\Phi}_\text{ansatz})$ is also an entanglement measure in vortex systems.
Then we have
\begin{align}
    C({\Phi}_\text{ansatz})= \sqrt{4N^2\lambda^2(1-\lambda)^2cc^{\prime}}.
\end{align}
Clearly, $C({\Phi}_\text{ansatz})$ vanishes if the ansatz is non-entangled, i.e., $\lambda=0$.
In the simplest case that $X_1 =Y_1=X_2=Y_2=0$, $C({\Phi}_\text{ansatz}) = \Upsilon/\Lambda = 2\lambda(1-\lambda)/[\lambda^2+(1-\lambda)^2] \leq 1$, with equality if $\lambda=1/2$.
Note here that $C({\Phi}_\text{ansatz})$ is a monotonic and strictly increasing function.
Also, when $X_1 =X$, $Y_1=Y$, and $X_2=Y_2=0$, then we can show that $C({\Phi}_\text{ansatz}) \leq 1$, with equality if $\lambda=1/2$ and $X^2+Y^2=0$.
Interestingly, we find a functional relation between the concurrence and the von Neumann entanglement entropy
\begin{align}
    E({\Phi}_\text{ansatz}) = h\left(  \frac{1+\sqrt{1-C({\Phi}_\text{ansatz})^2}}{2}    \right),
\end{align}
where $h(x)=-x\log{x}- (1-x)\log (1-x)$ is the binary entropy function.

Let us generalize this functional relation by introducing the quantum Rényi-$\alpha$ entropy \cite{Adesso, Kim, Adesso2}
\begin{align}
    S_{\alpha}(\rho) = \frac{1}{1-\alpha} \log[\mathrm{tr}(\rho^\alpha)],
\end{align}
for any $\alpha>0$.
In the limit $\alpha \to 1$, $S_{\alpha}(\rho)$ converges to the von Neumann entropy, i.e., $\lim_{\alpha \to 1} S_{\alpha}(\rho) = S(\rho)$.
For a pure two-particle state $\Ket{\psi}_{AB}$, the Rényi-$\alpha$ entanglement entropy is defined by
\begin{align}
    E_\alpha(\psi_{AB}) = S_{\alpha}(\rho_A) = S_{\alpha}(\rho_B).
\end{align}
Here, $E_\alpha(\psi_{AB})$ vanishes on non-entangled states and is invariant under local unitary transformations.
For the ansatz, the Rényi-$\alpha$ entanglement entropy is given by
\begin{align}
    E_\alpha({\Phi}_\text{ansatz}) = \frac{1}{1-\alpha} \log[p_+^\alpha + p_-^\alpha].
\end{align}
Then we obtain the generalized functional relation between the concurrence and the Rényi-$\alpha$ entanglement entropy
\begin{align}
    E_\alpha({\Phi}_\text{ansatz}) = f_\alpha\left(C({\Phi}_\text{ansatz}) \right),
\end{align}
where for $\alpha>0$
\begin{align}
    f_\alpha(x)=\frac{1}{1-\alpha}\log\left[\left(\frac{1+\sqrt{1-x^2}}{2}\right)^\alpha +\left(\frac{1-\sqrt{1-x^2}}{2}\right)^\alpha\right].
\end{align}
In the limit $\alpha \to 1$, the function $f_{\alpha}(x)$ converges to the  binary entropy function, i.e., $\lim_{\alpha \to 1} f_{\alpha}(x) = h(x)$.

%%%%%%%%%%%%%%%%%%%%%%%%%%%%%%%%%%%%%%%%%%%%%%%%%%%%%%%%%%%%%%%%%%%%%%%%%%%%%%%%%%%%%%%%%%%%%%%%%%%%%%%%%%%%%%%%%%%
\subsection*{6. Dirac's generalized canonical formulation}
Let us develop the framework of the vortex dynamics based on Dirac's generalized canonical formulation.
As explicitly as possible, we fix the vortex coordinate $(X_2,Y_2 )$ at the origin $(0,0)$.
The reduced Lagrangian (\ref{L2}) thus takes a simple form
\begin{align} \label{TL}
L = &\frac{\alpha E({\dot{X}}Y - X{\dot{Y}})}{\Lambda[1+\alpha(X^2+Y^2)]}-\frac{\alpha}{2[1+\alpha(X^2+Y^2)]},
\end{align}
where we take out the notation of subscript of $(X_1,Y_1 )$.
Moreover, let us perform the point transformation from $(X,Y)$ to dimensionless variables $(\xi, \eta)$
\begin{align} \label{T1}
\xi  &=X \sqrt{ \frac{2\alpha E} {\Lambda[1+\alpha(X^2+Y^2)]}},
\\       \label{T2}
\eta&=Y \sqrt{ \frac{2\alpha E} {\Lambda[1+\alpha(X^2+Y^2)]}}.
\end{align}
The reduced Lagrangian is thus rewritten as
\begin{align} \label{CTL}    
L= \frac {\dot{\xi}\eta -\xi\dot{\eta}} {2} +\frac {\alpha \Lambda} {4E} ({\xi}^2 + {\eta}^2 ),
\end{align}
where an irrelevant additive constant has been included.
Here, without loss of generality, we can determine the signs of the pointlike vortices, such as $\epsilon_1 = -1$, $\epsilon_2=+1$, and therefore $E=1-2\lambda >0$.
This point transformation is regular since the Jacobian determinant does not vanish:
\begin{align}
    \frac{\partial (\xi,\eta)}{\partial (X,Y)} =\frac{2\alpha E}{\Lambda[1+\alpha(X^2+Y^2)]^2}   \neq 0.
\end{align}
According to the identical equation
\begin{align}
    1+\alpha(X^2+Y^2) = \frac{2E}{2E-\Lambda({\xi}^2 + {\eta}^2)},
\end{align}
we have
\begin{align}
    {\xi}^2 + {\eta}^2 < \frac{2E}{\Lambda}.
\end{align}

Let us apply this reduced Lagrangian formulation to the Dirac's generalized canonical formalism \cite{Dirac01}.
We can then show the canonical conjugate nature of the canonical coordinates $\xi$ and $\eta$ (see Appendix):
\begin{align} \label{Dirac2ss}
\{\xi, \eta\}_D = 1,
\end{align}
where $\{, \}_{D}$ is the Dirac bracket.
The relation (\ref{Dirac2ss}) is in analogy with the one for the pointlike vortex system in classical $2$-dimensional flow \cite {classicallamb32}.
With the help of the Legendre transformation, the reduced Hamiltonian is written as
\begin{align} \label{Ha1}
H &= \dot{\xi} p_\xi + \dot{\eta} p_\eta -L 
\nonumber\\
&=- \frac {\alpha \Lambda} {4E} ({\xi}^2 + {\eta}^2 ).
\end{align}
Since the Hamilton's equations are written by $\dot{f} = \{f,H\}_D$, we have
\begin{align} \label{EL3}
\dot{\xi} &=-\kappa \alpha \eta,
\\ \label{EL4}
\dot{\eta} &= \kappa \alpha \xi,
\end{align}
where $\kappa$ is a positive constant given by
\begin{align}
    \kappa =\frac{\lambda}{2E}.
\end{align}
It is evident that Eqs.~(\ref{EL3}) and (\ref{EL4}) are the same as the Euler-Lagrange equations calculated from the reduced Lagrangian (\ref{CTL}).

Let us compare our results with the classical two-dimensional flow.
Clearly, $\xi^2 + \eta^2$ is a constant of motion.
This reminds us of the Hamiltonian-like formalism of classical vortex systems in two dimensions \cite {classicallamb32}.
In classical hydrodynamics, the Hamiltonian-like function is in the logarithm form of the constant of motion, while the present reduced Hamiltonian is not the case.

Finally, we analyze the vortex dynamics.
Calculating the Euler-Lagrange equations for the vortex variables in the reduced Lagrangian (\ref{TL}), we find that the vortex coordinates behave like the harmonic oscillator:
\begin{align}
    \ddot{X}(t) + \omega^2 X(t) &= 0, \\
    \ddot{Y}(t) + \omega^2 Y(t) &= 0,
\end{align}
where the angular frequency $\omega$ is defined by
\begin{align}
    \omega &= \kappa \alpha \nonumber\\
    &=\frac{\lambda^2+(1-\lambda)^2}{2(1-2\lambda)} \alpha,
\end{align}
and $X^2 + Y^2$ is a constant of motion.
An interesting situation arises when the entanglement parameter $\lambda$ changes.
If $\lambda = 0$, then the angular frequency $\omega$ is equal to $\alpha/2$.
On the other hand, if $\lambda \neq 0$, then $\omega$ monotonically increases as $\lambda$ approaches to $1/2$; in the limit $\lambda \to 1/2$, $\omega$ goes to infinity.
This implies that entanglement effects are also linked to the rotational speed up in vortex systems.
That is, the stronger entanglement is, the faster the dynamics become.
Following the nonexistence of entanglement in classical hydrodynamics, we can conclude that this result goes beyond straightforward analogies and provides a characteristic feature of the quantum formalism.

%%%%%%%%%%%%%%%%%%%%%%%%%%%%%%%%%%%%%%%%%%%%%%%%%%%%%%%%%%%%%%%%%%%%%%%%%%%%%%%%%%%%%%%%%%%%%%%%%%%%%%%%%%%%%%%%%%%%%%%%%%%%%%%%%%%%%%%%%%%%%%%%%%%%%%%%%%%%%%%%%%%%%%%%%%%%%%%%%%%%%%%%%%%%%%%%%%%%%%%%%%%%

\subsection*{7. Conclusion}
In this paper, we have considered the hydrodynamic representation of quantum mechanics for two particles with nonintegrable phases.
We have introduced the ansatz for the entangled wavefunction representing pointlike vortices.
Based on the time-dependent variational principle for the Schr\"{o}dinger equation combined with the Rayleigh-Ritz method, we have obtained the reduced Lagrangian for the vortex variables.
Then we have discussed the effects of entanglement on the approximate vortex-vortex dynamics.
By considering some limitations, we have shown that entanglement can strongly complicate vortex dynamics.
This result can contain a seed of potentially interesting issues from mathematical and physical viewpoints.
Moreover, we have also found that strong entanglement can be linked to the rotational speed up in vortex systems.
This result can go beyond simple analogies with classical vortex dynamics.
Also, we have provided a quantitative analysis of entanglement in vortex systems.
Finally, we have developed the formulation of vortex dynamics using Dirac's generalized Hamiltonian formalism.

In quantum information theory, many works study how quantum dynamics (or operations) affect entanglement.
On the other hand, we have advanced the converse question of how entanglement affects dynamics.
Exploring this direction might yield new insights into aspects of entanglement.
It also would be interesting to apply our special non-Gaussian entangled wavefunctions with nonintegrable phases in vortex systems to quantum information processing tasks.

%%%%%%%%%%%%%%%%%%%%%%%%%%%%%%%%%%%%%%%%%%%%%%%%%%%%%%%%%%%%%%%%%%%%%%%%%%%%%%%%%%%%%%%%%%%%%%%%%%%%%%%%%%%%%%%%%%%%%%%%%%%%%%%%%%%%%%%%%%%%%%%%%%%%%%%%%%%%%%%%%%%%%%%%%%%%%%%%%%%%%%%%%%%%%%%%%%%%%%%%%%%%

\subsection*{Acknowledgments}
I would like to thank Otfried G\"{u}hne for careful reading of the manuscript.
This work has been supported by the DFG and the ERC (Consolidator Grant 683107/TempoQ).

%%%%%%%%%%%%%%%%%%%%%%%%%%%%%%%%%%%%%%%%%%%%%%%%%%%%%%%%%%%%%%%%%%%%%%%%%%%%%%%%%%%%%%%%%%%%%%%%%%%%%%%%%%%%%%%%%%%%%%%%%%%%%%%%%%%%%%%%%%%%%%%%%%%%%%%%%%%%%%%%%%%%%%%%%%%%%%%%%%%%%%%%%%%%%%%%%%%%%%%%%%%%
\subsection*{Appendix: Canonical conjugate nature of $\xi$ and $\eta$}
Here, we show the canonical conjugate nature of the canonical coordinates $\xi$ and $\eta$.
We begin by considering the canonical momenta, $p_\xi = {\partial L}/{\partial \dot{\xi}}={\eta}/{2}$, and $p_\eta = {\partial L}/{\partial \dot {\eta}}=-{\xi}/{2}$.
These lead to the weak constraints: $\chi_\xi = p_\xi -{\eta}/{2} \approx 0$, $\chi_\eta = p_\eta + {\xi}/{2} \approx 0$,
where the symbol ``$\approx$'' is called the weak equality.
It is considered that the weak constraints are the inner ones coming from the structure of the reduced Lagrangian itself, and, in principle, the dynamics of the constrained vortex systems can be described by the two canonical variables $\xi$, $\eta$.

The Poisson bracket is written as
\begin{align}
\{A,B\}_P = \sum_{\zeta =\xi,\eta} \left( \frac{\partial A}{\partial \zeta} \frac{\partial B}{\partial p_\zeta}
                                                    -\frac{\partial A}{\partial p_\zeta }   \frac{\partial B}{\partial \zeta}  \right),
\end{align} 
where $A$ and $B$ are the functions of $(\xi,p_\xi,\eta,p_\eta)$.
Therefore, we have that $\{\chi_\xi,\chi_\eta \}_P = -1$, which does not vanish.
We propose that $\chi_\xi$ and $\chi_\eta$ are the second-class constraints in Dirac's generalized canonical formalism \cite{Dirac01}.
In the present case, it is standard to apply not the Poisson bracket but the Dirac bracket defined by
\begin{align} \label{Dirac1}
\{A,B\}_D = \{A,B\}_P - \sum_{a,b = \xi,\eta} \{ A, \chi_a\}_P \,{C_{ab}^{-1}} \, \{ \chi_b, B\}_P.
\end{align}
Here, $ C_{ab}$'s are matrix elements satisfying $C_{ab} = \{\chi_a,\chi_b\}_P$.
That is, $C_{\xi \xi} = -C_{\eta \eta}=0$, and $C_{\xi \eta} = -C_{\eta \xi}=-1$.
Note that the second-class constraints are the identical equations in the Dirac bracket.
Hence, we have 
\begin{align} \label{Dirac2}
\{\xi, \eta\}_D = 1.
\end{align}

%%%%%%%%%%%%%%%%%%%%%%%%%%%%%%%%%%%%%%%%%%%%%%%%%%%%%%%%%%%%%%%%%%%%%%%%%%%%%%%%%%%%%%%%%%%%%%%%%%%%%%%%%%%%%%%%%%%%%%%%%%%%%%%%%%%%%%%%%%%%%%%%%%%%%%%%%%%%%%%%%%%%%%%%%%%%%%%%%%%%%%%%%%%%%%%%%%%%%%%%%%%%

\end{document}